\documentclass[aps,preprint,showpacs,superscriptaddress,groupedaddress]{revtex4}  
\usepackage{graphicx}  
\usepackage{dcolumn}   
\usepackage{bm}        
\usepackage{amssymb}   
\usepackage{amsmath}
\usepackage{makecell}
\usepackage{scrextend}

\begin{document}

\title{Non-linear alloying and strain effects on trivial-topological and semimetal-semiconductor transitions in Bi$_{1-x}$Sb$_x$}
\def\ITA{Instituto Tecnol\'{o}gico de Aeron\'{a}utica, 12228-900 S\~{a}o Jos\'{e} dos Campos, SP, Brazil}
\def\JENA{Institut f\"ur Festk\"orpertheorie und -optik, Friedrich-Schiller-Universit\"at, 07743 Jena, Germany}

\author{F.L.~Freitas} \email{felipelopesfreitas@gmail.com} \affiliation{\ITA,\JENA}
\date{\today}

\begin{abstract}
Applying the approximate DFT-1/2 quasiparticle scheme, band structure unfolding, and generalized quasichemical approximation to describe chemical and structural disorder, we investigate the electronic structure of Bi$_{1-x}$Sb$_x$ alloys from first principles. We calculate the important energy levels near the Fermi energy versus the Sb concentration $x$ where the trivial-topological (TT) and semimetal-semiconductor (SMSC) transitions occur. We demonstrate that the energy variation of the relevant states deviates significantly from linear behavior and that the bowings are important to correctly describe the critical compositions. The influence of strain on the energy levels is briefly discussed. It is concluded that the type or sign of strain applied on antimony atoms during the growth of the alloy should be heavily dependent on its composition.
\end{abstract}

\maketitle

\section{\label{sec:intro}Introduction}

Bismuth, antimony and their alloys (Bi$_{1-x}$Sb$_x$) are materials that have been subject of extensive studies because of their unusual properties among semimetals and narrow band-gap semiconductors. In the semiconducting phase, Bi$_{1-x}$Sb$_x$ alloys possess a very high thermoelectric efficiency illustrated by a huge figure of merit at temperatures around 80 K, making them desirable for applications as thermocouples and thermoelectric coolers\cite{Smith1962,Lin2001}.

In addition to that, bismuth antimony alloys have several peculiar characteristics, which come about as a consequence of the drastic dependence of the valence and conduction bands on many controllable physical parameters. Varying the alloy composition, the semimetallic character of pure Bi and Sb with inverted band structure changes into semiconducting phases for intermediate compositions of Bi$_{1-x}$Sb$_x$. Thereby, the very strong spin-orbit interaction (SOI) in Bi and Sb atoms gives rise to interesting topological properties. The topology of pure Sb (Bi) is characterized by the three-dimensional $\mathbb{Z}_2$ class of invariants $(1;111)$ [$(0;000)$]\cite{Fu2007a,Kane2005}. The alloys with $x<0.22$ have been identified as the first non-Quantum-Hall topological insulators (TIs). Indeed, massless Dirac particles have been observed by angle-resolved photoemission spectroscopy (ARPES) in the gapless topological surface bands\cite{Hsieh2008}. Theoretical predictions indicate that they should give rise to a novel phase of matter analogous to the Quantum Spin Hall (QSH) effect known in two-dimensional TIs\cite{Fu2007b}. The experimental observation of these topological states in Bi$_{1-x}$Sb$_x$\cite{Hsieh2008} suggests this material could play a pivotal role in spintronics\cite{Wang2016} and quantum computing\cite{Qi2009,Hasan2010}, areas where topological insulators are expected to have groundbreaking applications.

Depending on the Sb concentration, the Bi$_{1-x}$Sb$_x$ alloy system can be either a semiconductor or a semimetal, and the valence band might have a different character, making the system trivial or topological. Mainly from (magneto)transport measurements there is experimental consensus about the critical compositions $x$ of the semimetal-semiconductor transitions (see Ref. \onlinecite{Cho1999} and references therein) and Lifschitz electronic-topological transitions (see Ref.\onlinecite{Kosarev2007} and references therein). Theoretical band structure studies\cite{Fu2007b,Hsieh2008} qualitatively agree but suffer from the limitations of the used tight-binding (TB) approach\cite{Liu1995} and the virtual crystal approximation (VCA) to interpolate between the end components Bi and Sb. The VCA ignores disorder due to the random mixture of Bi and Sb atoms. The phenomenological TB model does not really agree with experimental results\cite{Hsieh2008} as explicitly shown for the topological surface states\cite{Teo2008}. First-principles electronic structure calculations in the framework of the Density Functional Theory (DFT) in local density approximation\cite{Liu2011} suffer from the lack of quasiparticle (QP) effects and, hence, cannot give trustable band orderings.

Moreover, recently there was progress to grow Bi$_{1-x}$Sb$_x$ thin films on several substrates with different epitaxy techniques (see e.g. Ref. \onlinecite{Cho1999}). The pseudomorphic growth induces biaxial strain in the overlayer, which significantly shifts the band extrema. However, the influence of strain on trivial-topological as well as semimetal-semiconductor phase transitions has been only discussed for pure Bi\cite{Aguilera2015} and is unknown for alloys.

The characterization of the electronic and topological properties of strained or unstrained Bi$_{1-x}$Sb$_x$ as a TI demands a precise analysis of how the energies of various states depend on Sb composition $x$. Therefore, in this article we investigate the electronic structure and the phase transitions in Bi$_{1-x}$Sb$_x$ from first principles including QP effects, spin-orbit coupling, the random distribution of Bi and Sb, and strain.

The paper is organized as follows: in Sec. II a description of the computational method is given, whereas the detailed discussion of the results of the simulation is provided in Sec. III. Finally, all conclusions are summarized in Sec. IV.

\section{Calculational methods}

\subsection{Crystal structure of Bi and Sb and cluster expansion}

The description of the electronic properties of the Bi$_{1-x}$Sb$_x$ system starts with an analysis of the crystal structure of the pure materials Bi and Sb. It is well known that both bismuth and antimony crystallize in the rhombohedral A7 structure\cite{Falicov1965,Falicov1966} with two atoms in the unit cell, which can be visualized as two face-centered cubic networks of atoms deformed by a shear angle.

For the purposes of this work, it is better to represent the crystal as a hexagonal supercell (see Ref. \onlinecite{Aguilera2015}), containing six atoms, and whose geometry is defined by the hexagonal lattice parameters $a$ and $c$, and another parameter $\mu$ describing the separation between the two basis atoms\cite{Liu1995}.

This supercell description is particularly useful for the method chosen to describe the properties of the alloy Bi$_{1-x}$Sb$_x$, the Generalized Quasichemical Approximation (GQCA)\cite{Sher1987}, discussed in the context of \emph{ab initio} simulations both for binary\cite{Freitas2016a} and (pseudo)ternary\cite{Freitas2016b} in previous studies. For completeness of the present work, we discuss the approximation here briefly.

Within the GQCA, the alloy is treated as a sum of clusters, all with the same total number of atoms $n$, which are assumed to be energetically independent. The possible types of clusters are, of course, limited in number. For a binary alloy like Bi$_{1-x}$Sb$_x$, simple combinatorial arguments give a total number of $2^n$ possible cluster configurations. If the size of these clusters is very large, one can treat any macroscopic property $P$ of the statistical system at a given composition $x$ as a mean of the property in each cluster configuration $P_j$ averaged by the probability of occurence of the $j$-th configuration $x_j(x)$:

\begin{equation}
P(x) = \sum_j x_j(x)P_j.
\end{equation}

The most remarkable feature of the GQCA is that the probabilities $x_j(x)$ are truly random only at the limit where the growth temperature $T\to\infty$. For finite $T$ (which is around 100$^\circ$ C for Bi$_{1-x}$Sb$_x$\cite{Cho1999}), it is possible to calculate deviations from random behavior in function of the formation energy per atom of each cluster configuration $\epsilon_j$. Since crystallographic structures are highly symmetric, it is expected that many clusters will share identical formation energies, and can be grouped into classes with a degeneracy factor $g_j$. By writing $\beta = 1/kT$, the probability of each cluster class is given by

\begin{equation}
x_j(x) = \frac{g_j\eta(x)^{n_j}e^{-\beta\epsilon_j}}{\sum_j g_j \eta(x)^{n_j}e^{-\beta\epsilon_j}},
\label{eq:prob}
\end{equation}
where $n_j$ is the number of atomic species of a given type, and $\eta$ is a positive adimensional parameter obtained by imposing the constraint

\begin{equation}
\sum_j x_j(x) n_j = nx,
\end{equation}
which gives an $n$-th degree polynomial equation for $\eta$, easily solved by standard numerical methods. For the system discussed in this work, it was verified that for the entire range of compositions $x$ in the Bi$_{1-x}$Sb$_x$ alloy, the polynomial equation admits only one real positive root, and thus there is only one possible phase.

The quality of the GQCA is dictated by the size of the cluster. In practice, when combining it with first-principles methods, the size is dictated mostly by the computational capabilities of the hardware running the simulation. For the present work, the clusters considered were the six-atom hexagonal supercells mentioned previously, for a total of $2^6=64$ possible configurations.

The cluster configurations were grouped into 13 distinct cluster classes by group-theoretical considerations. The crystal structure A7 has space group $R\bar{3}m$, with point group symmetry $\bar{3}m$, corresponding to 12 symmetry operations. By assigning labels to each atom in the supercell, the application of a symmetry operation will induce a permutation of the atomic labels.

The action of the point group on the hexagonal six-atom supercell is, thus, isomorphic to a group of permutations. Like any group, it should admit some primitive operations as generators\cite{Hungerford2003}. An hexagonal crystal structure can be seen as layers of atoms stacked in an A-B-C manner\cite{Ashcroft1976}. In the A7 structure, two distinct hexagonal networks are identified, separated by a fractional distance of $2\mu$\cite{Liu1995}. We can, therefore, label all sites in the supercell by their position in the A-B-C stack and which network they belong to. It is, then, possible to prove that the three primitive operations provided in Table~\ref{tab:op} generate all permutations corresponding to the 12 symmetry operations.

\begin{table}[h]
\caption{\label{tab:op}Transformations that generate the 12 permutations corresponding to the symmetry operations of the A7 crystal structure. The labelling of each atomic site is explained in the text.}
\begin{ruledtabular}
\begin{tabular}{cccc}
Op. 1 & \thead{$A1\to B1$ \\ $A2\to B2$} & \thead{$B1\to C1$ \\ $B2\to C2$} & \thead{$C1\to A1$ \\
$C2\to A2$} \\
\hline
Op. 2 & \thead{$A1\to A1$ \\ $A2\to A2$} & \thead{$B1\to C1$ \\ $B2\to C2$} & \thead{$C1\to B1$ \\
$C2\to B2$} \\
\hline
Op. 3 & \thead{$A1\to A2$ \\ $A2\to A1$} & \thead{$B1\to B2$ \\ $B2\to B1$} & \thead{$C1\to C2$ \\
$C2\to C1$} \\
\end{tabular}
\end{ruledtabular}
\end{table}

In possession of these three primitive operations, a simple algorithm can group each cluster configuration in an inequivalent class. First, one generates all $2^6=64$ possible configurations by placing either a Bi or Sb atom in the A1 to C2 atomic sites. Then, one applies each of the three operations in Table~\ref{tab:op}, generating a new configuration. After this procedure, a graph (in the computer science definition) is obtained, and by using a Depth-First Search algorithm\cite{Cormen2001}, one finds 13 distinct connected components, whose size gives the degeneracy factor in Eq.~\eqref{eq:prob}. Any of the configurations within a connected component can be used for the simulation.

\begin{table}
\caption{\label{tab:class}Number of atoms of each species, together with their symmetry factor, in each of the cluster classes used for the calculation of the Bi$_{1-x}$Sb$_x$ alloy.}
\begin{ruledtabular}
\begin{tabular}{cccc}
No. & Bi atoms & Sb atoms & Symmetry \\
\hline
1 & 6 & 0 & 1 \\
2 & 5 & 1 & 6 \\
3 & 4 & 2 & 6 \\
4 & 3 & 3 & 2 \\
5 & 4 & 2 & 3 \\
6 & 4 & 2 & 6 \\
7 & 3 & 3 & 12 \\
8 & 3 & 3 & 6 \\
9 & 2 & 4 & 6 \\
10 & 2 & 4 & 3 \\
11 & 2 & 4 & 6 \\
12 & 1 & 5 & 6 \\
13 & 0 & 6 & 1 \\
\end{tabular}
\end{ruledtabular}
\end{table}

Each of these 13 cluster classes is presented in Table~\ref{tab:class}, one notices that the number of configurations in each class is always a divisor of the total number of symmetry operations (12), as expected from group theory\cite{Hungerford2003}.

\subsection{Band structures of Bi and Sb with quasiparticle corrections}

The next step in the accurate determination of the topological and electronic properties of the alloy is the appropriate description of the bandstructures of the pure materials Bi and Sb. This presents two challenges. First, it is necessary to determine the structural properties of the system very accurately (i.e., the lattice parameters must not deviate much from experimentally measured values). This happens because for Bi and Sb, the band gap is of the order of 0.1 eV, and even small imprecisions in the lattice calculation lead to highly unphysical properties. In our tests, we found out that the PBEsol exchange-correlation functional\cite{Perdew2008} provides the most accurate structural parameters, presented in Table~\ref{tab:latpars}. 

\begin{table}[h]
\caption{\label{tab:latpars}Lattice parameters (in \AA) calculated in PBEsol, compared to measured values, and L point energy gaps (in meV) obtained within DFT and DFT-1/2, also compared to experimental results, for bismuth and antimony.}
\begin{ruledtabular}
\begin{tabular}{cccccc}
& $a$(th./exp.) & $c$(th./exp.) & $E_{gL}^{DFT}$ & $E_{gL}^{-1/2}$ & $E_{gL}^{exp}$\\
\hline
Bi & 4.496/4.533\footnote{\label{foot:lat}Ref.~\onlinecite{Schiferl1969}} & 11.735/11.797\footref{foot:lat} & 105 & 20 & 13\footnote{Ref.~\onlinecite{Vecchi1974}.}\\
Sb & 4.309/4.301\footref{foot:lat} & 11.248/11.222\footref{foot:lat} & 87 & 237 & 219\footnote{Ref.~\onlinecite{Brandt1982}, estimate.}\\
\end{tabular}
\end{ruledtabular}
\end{table}

Second, it is well known that conventional \emph{ab initio} calculations based on Density Functional Theory (DFT)\cite{Kohn1965} are plagued by the so-called \emph{band-gap problem}. Since the eigenvalues of the Kohn-Sham equation represent energies of a fictitious non-interacting system, the energy gap derived from them is usually much smaller than the real one.

The standard way to fix this problem is to go beyond DFT and apply the GW approximation\cite{Aryasetiawan1998}, derived within Many-Body theory. Indeed, it has been reported that, despite some complications due to Spin-Orbit Coupling (SOC) effects, GW calculations improve upon plain DFT\cite{Aguilera2015}, although don't really reach the required meV accuracy. In addition to that, it has a very serious shortcoming. The incorporation of the quasiparticle effects within GW requires the solution of an integro-differential equation, and is very computationally demanding, being unsuitable for the calculation of even relatively small systems like the six-atom clusters considered in the present work.

In order to accurately determine the band structures, we use the DFT-1/2 method\cite{Ferreira2008}, which presents an accuracy similar to GW in many semiconductors, and has been applied successfully to the study of alloys\cite{Freitas2016a,Freitas2016b}. In the DFT-1/2 framework, the band gap is corrected by removing half electron from the atomic orbital corresponding to the top of the valence band. For group-IV semiconductors like Si\cite{Ferreira2008}, Ge and Sn\cite{Freitas2016a,Freitas2016b}, this half electron is effectively split among the two atoms in the unit cell. In the group-V semimetals Bi and Sb, a similar phenomenon happens, the main difference is that the atomic \emph{p} orbital of the atoms only contributes 80\% to the valence band, and thus $0.20$ electron must be removed from the DFT calculation.

However, the charge removal causes divergences in the calculations, due to the long-range nature ($\sim 1/r$) of the Coulomb potential. For that reason, one imposes a cutoff, defined by the CUT parameter, chosen variationally to extremize the direct energy gap at the $L$ point of the Brillouin zone (BZ). The $L$ point is extremely important in the determination of the topological character of the system, as will become clear in the next sections.

\begin{figure}[h!]
\includegraphics{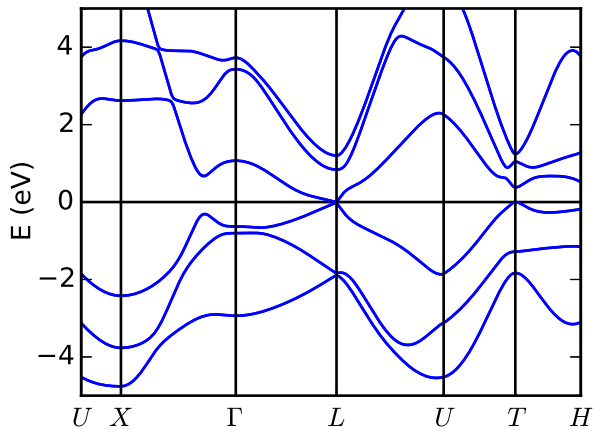}
\caption{\label{fig:biband} Calculated DFT-1/2 band structure for bismuth. The reference energy is at the top of the valence band.}
\end{figure}

\begin{figure}[h!]
\includegraphics{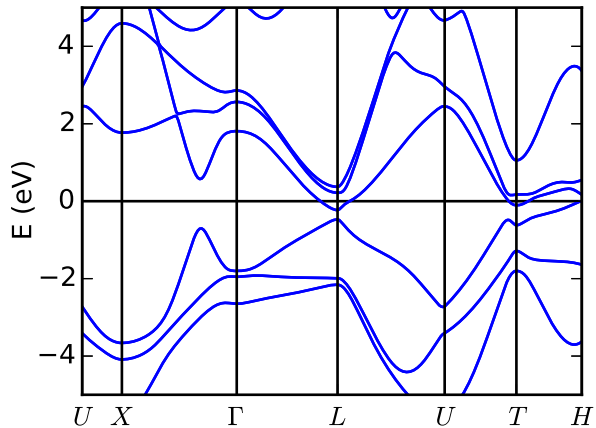}
\caption{\label{fig:sbband} Calculated DFT-1/2 band structure for antimony. The reference energy is at the top of the valence band.}
\end{figure}

The obtained band structures for both Bi and Sb are provided in Figs.~\ref{fig:biband} and~\ref{fig:sbband}. The DFT simulation was carried out by approximating the all-electron wavefunctions and pseudopotentials in the Projector Augmented Wave (PAW) scheme\cite{Blochl1994} as implemented in the Vienna \emph{ab initio} Simulation Package (VASP)\cite{Kresse1996a,Kresse1996b}. A cutoff of 520 eV was chosen for the plane-wave expansion, and atomic positions were relaxed until all Hellman-Feynman forces reached a magnitude smaller than 0.01 eV/\AA. The BZ integration was approximated by a discrete sampling on a $6\times6\times2$ $\Gamma$-centered Monkhorst-Pack\cite{Monkhorst1976} grid. In each alloy cluster, the DFT-1/2 method was implemented by correcting the atomic pseudopotentials with the removal of $0.20$ electron from the $p$ orbital of the Bi and Sb atoms, in both cases with CUT=4.1 a.u.

The accuracy of the obtained band gap energies at the $L$ point, $E_{gL}$, in comparison to plain DFT and experimental values, is shown on Table~\ref{tab:latpars}. Curiously, the DFT calculation \emph{overestimates} the gap at Sb, this behavior is typical for inverted band gaps like that of Sb and has been observed in other systems applying various quasiparticle correction methods\cite{Yazyev2012,Kufner2015}.

\subsection{Topological character of Bi$_{1-x}$Sb$_x$ binary alloy}

The analysis of the topological properties of a general condensed matter system is normally complicated, especially in first-principles simulations. As described in Ref.~\onlinecite{Fu2007a}, in three dimensions, periodic systems can be grouped in 16 distinct classes, determined by four $\mathbb{Z}_2$ invariants $(\nu_0;\nu_1\nu_2\nu_3)$. However, most methods to determine these invariants require that the phases of the wavefunctions satisfy some specified conditions (gauge fixing), which does not happen in \emph{ab initio} calculations, where the phases are random.

Although there are methods which can be applied without fixing the gauge\cite{Yu2011}, it has been noted that, for systems with parity inversion\cite{Fu2007b,Fruchart2013} the calculation simplifies enormously, and it is only necessary to look at the parity of the wavefunctions of Time-Reversal Invariant Momenta (TRIM). For the case of bismuth and antimony, it has been noted\cite{Fu2007b} that such analysis of the TRIM points $\Gamma$, $X$, $L$ and $T$ reveals that on Bi, there are two inversions, at $L$ and $T$, giving a trivial topology, while Sb only shows one inversion at $T$, and has nontrivial topology.

However, neither Bi or Sb are insulators, since their conduction band minimum (CBM) lies below the valence band maximum (VBM), which makes them semimetals. For bismuth, the CBM and VBM are, respectively, at the $L$ and $T$ points, while in antimony they are at the $L$ and $H$ points. The $H$ point is a six-fold degenerate point of the BZ, whose precise coordinates depend on the band structure calculation. In this work, we determine its trigonal coordinates at [0.3601,0.1563,0.1563].

The fact that the VBM is at two distinct points on Bi and Sb allows us to drastically change the character of the system by alloying. By mixing the two materials, it is expected that, at some intermediate concentration, both $T$ and $H$ have a sufficiently low energy in order to open a gap and turn the system into a semiconductor, giving rise to two semimetal-semiconductor (SMSC) transitions. In addition to that, we also expect a change on the inverted character of the bands at the $L$ point, where the system changes its topological character.

More precisely, the topological character is determined by the parity of the conduction and valence states at the $L$ point. In Bi, the antissymetric (odd) $L_a$ state is below the symmetric (even) $L_s$ state, while in Sb it is above. Therefore, at a given Sb composition $x$, there is a trivial-topological (TT) transition. When, in addition to that, the system possesses a gap, we expect the Bi$_{1-x}$Sb$_x$ alloy to be a strong topological insulator.

It should be clear that, in order to predict theoretically when the alloy is TI, it is necessary to trace the evolution of the energies of the states $L_a$, $L_s$, $T$ and $H$ across the entire composition range. This presents three challenges to our approach.

The first one, treated in this section, concerns on how we can identify these states in a given cluster. It is well-known\cite{Freitas2016b} that, when performing calculations on supercells, the states defined on the unit cell get reflected to other points of the BZ of the supercell, and it becomes necessary to define a systematic way to unfold the bands and retrieve them.

It turns out that the band unfolding operation can be described in a very simple manner using matrices. In our approach, we use a six-atom supercell whose primitive lattice vectors $\mathbf{c}_i$ depend on the lattice vectors $\mathbf{a}_i$ of the unit cell as

\begin{equation}
\begin{bmatrix}
\mathbf{c}_1 \\ \mathbf{c}_2 \\ \mathbf{c}_3
\end{bmatrix} =
\begin{bmatrix}
-1 & 1 & 0 \\
0 & -1 & 1 \\
1 & 1 & 1
\end{bmatrix}
\begin{bmatrix}
\mathbf{a}_1 \\ \mathbf{a}_2 \\ \mathbf{a}_3
\end{bmatrix}.
\end{equation}

This dependence can be written as a matrix equation $c=Ma$. Denoting by $\mathbf{b}_i$ the reciprocal lattice vectors of the unit cell and by $\mathbf{d}_i$ those of the supercell, we must have\cite{Ashcroft1976}

\begin{equation}
\begin{bmatrix}
\mathbf{b}_1 \\ \mathbf{b}_2 \\ \mathbf{b}_3
\end{bmatrix}
\begin{bmatrix}
\mathbf{a}_1 & \mathbf{a}_2 & \mathbf{a}_3
\end{bmatrix} = 
\begin{bmatrix}
\mathbf{d}_1 \\ \mathbf{d}_2 \\ \mathbf{d}_3
\end{bmatrix}
\begin{bmatrix}
\mathbf{c}_1 & \mathbf{c}_2 & \mathbf{c}_3
\end{bmatrix} =
2\pi
\begin{bmatrix}
1 & 0 & 0 \\
0 & 1 & 0 \\
0 & 0 & 1 \\
\end{bmatrix}.
\end{equation}

Therefore, if we have $c=Ma$, we must have $d=(M^T)^{-1}b$, so that $d^Tc = b^T M^{-1}Ma = b^Ta = 2\pi I$. An arbitrary $k$-point can be written in function of the reciprocal vectors of the unit cell or the supercell as

\begin{equation}
\mathbf{k} =
\begin{bmatrix}
v_1 & v_2 & v_3
\end{bmatrix}
\begin{bmatrix}
\mathbf{b}_1 \\ \mathbf{b}_2 \\ \mathbf{b}_3
\end{bmatrix} =
\begin{bmatrix}
u_1 & u_2 & u_3
\end{bmatrix}
\begin{bmatrix}
\mathbf{d}_1 \\ \mathbf{d}_2 \\ \mathbf{d}_3
\end{bmatrix}.
\end{equation}

In order for the $\mathbf{k}$ vector to remain invariant, we must have $u=Mv$, because this implies $u^Td = v^TM^Td = v^TM^T(M^T)^{-1}b = v^Tb$. Therefore, to find a given $k$-point in the supercell, one only needs to multiply its coordinates in the unit cell by the matrix

\begin{equation}
M = 
\begin{bmatrix}
-1 & 1 & 0 \\
0 & -1 & 1 \\
1 & 1 & 1
\end{bmatrix}.
\end{equation}

The supercell coordinates for all relevant k-points to this work are presented in Table~\ref{tab:coords}. This information allows us to identify unambiguously the $T$ and $H$ states in all clusters. However, both the $L_a$ and $L_s$ states are at the same k-point. In order to identify them, we can examine the effect of the quasiparticle correction at the band structure. If $L_s$ is above $L_a$, the bands are inverted, and by artificially increasing the strength of the DFT-1/2 correction, we expect the band gap at $L$ to become smaller. Similarly, if $L_s$ is below $L_a$, the band gap increases with stronger quasiparticle corrections. That allows us to unambiguously identify the $L_s$ and $L_a$ states in the clusters. It was verified that, in all 13 clusters, only the pure Bi one shows $L_s$ with greater energy than $L_a$.

\begin{table}
\caption{\label{tab:coords}Coordinates of the points $L$, $T$ and $H$ in the unit cell and the cluster supercell.}
\begin{ruledtabular}
\begin{tabular}{ccc}
Name & Unit cell & Supercell\\
\hline
L1 & [0.5,0.0,0.0] & [0.5,0.0,0.5] \\
L2 & [0.0,0.5,0.0] & [0.5,0.5,0.5] \\
L3 & [0.0,0.0,0.5] & [0.0,0.5,0.5] \\
T & [0.5,0.5,0.5] & [0.0,0.0,0.5]  \\
H1 & [0.3601,0.1563,0.1563] & [-0.2038,0,0.6727]  \\
H2 & [-0.3601,-0.1563,-0.1563] & [0.2038,0,-0.6727]  \\
H3 & [0.1563,0.3601,0.1563] & [0.2038,-0.2308,0.6727]  \\
H4 & [-0.1563,-0.3601,-0.1563] & [-0.2038,0.2308,-0.6727]  \\
H5 & [0.1563,0.1563,0.3601] & [0,0.2038,0.6727]  \\
H6 & [-0.1563,-0.1563,-0.3601] & [0,-0.2038,-0.6727]  \\
\end{tabular}
\end{ruledtabular}
\end{table}

\subsection{Definition of the quantum states in the statistical system}

The second challenge we face is to determine whether the energy of a given quantum mechanical state in a cluster calculation can be seen as a macroscopic property, and be averaged within the GQCA. The problem is exacerbated by the fact that \emph{ab initio} codes like VASP have different reference energies when calculating different systems, so it is not \emph{a priori} clear that we can simply take the energy of a state in different clusters and average it.

We now provide a proof that this can, in fact, be done. First, let us consider that we have calculated the average transition energy in the alloy, where the state's energy in the $j$-th cluster is taken with respect to the same reference level and denoted by $E_j$. The average energy is given by

\begin{equation}
E^{av}(x) = \sum_j x_j(x)E_j.
\end{equation}

In VASP, however, the reference energy is not constant, it is chosen so that the formation energy of an isolated atom is 0 eV\cite{VaspManual}. Therefore, we can understand the zero energy at a cluster calculation as the energy where all atoms are infinitely separated and thus, do not interact. In this limit, it is seen that the reference energy depends linearly on the number of atoms, and we have, in fact

\begin{equation}
E_j \to E_j + A + Bn_j,
\end{equation}
where $A$ and $B$ are constants which, as will be seen shortly, are unimportant. By averaging these energies, we obtain

\begin{equation}
E(x) = E^{av}(x) + A\sum_jx_j(x) + B\sum_jx_j(x)n_j.
\end{equation}

Both sums in the equation have fixed values, given the GQCA constraints $\sum_j x_j(x)=1$ and $\sum_j x_j(x)n_j = xn$. Since only differences in state energies at a given composition $x$ are observable, it is readily verified that by subtracting the energies of two given states computed in this manner, the result is the same as if the same reference energy were used:

\begin{equation}
E_2(x) - E_1(x) = E_2^{av}(x)-E_1^{av}(x).
\end{equation}

From this argument, it is clear that, at least for the convention for the reference energy used in VASP, it is perfectly fine to consider the state energy as a macroscopic property and perform an average as prescribed by the GQCA.

The third and final challenge is about whether it makes sense to define states such as $L_a$, $L_s$ and $H$ in a cluster with mixed atomic composition, or even in the alloy. The problem is a deep one, although there is experimental support for the argument that alloys possess an effective band structure, the theory behind it is still unclear\cite{Popescu2010}.

In the present work, this difficulty manifests itself in mixed cluster calculations, where the three-fold degenerate $L$ states and the six-fold degenerate $H$ states defined in the unit cell have their degeneracy lifted because of the different species of atoms in the supercell. During the simulations, it was verified that the differences in energies arising because of this symmetry breaking are of order 0.1 meV. Thus, it is possible to take their average energy to represent the quantum mechanical state in the statistical system, which is well defined on the scale of $\sim10$ meV needed to describe the topology of the system. 

\section{Results and discussion}

\begin{table}
\caption{\label{tab:calc}Formation energy per atom, and energies of the $L_a$, $L_s$, $T$ and $H$ states relevant to the calculation of the topological character of the Bi$_{1-x}$Sb$_x$ system. All values in eV. The reference energy is explained in the text.}
\begin{ruledtabular}
\begin{tabular}{cccccc}
No. & En./atom & $L_a$ & $L_s$ & $T$ & $H$ \\
\hline
1 & 0 & 6.2955 & 6.3187 & 6.3260 & 6.1600 \\
2 & 0.01081 & 6.3114 & 6.2155 & 6.2495 & 6.2313 \\
3 & -0.0095 & 6.2618 & 6.1061 & 6.1198 & 6.2430 \\
4 & -0.0540 & 6.4825 & 6.2947 & 6.0131 & 6.5689 \\
5 & 0.0289 & 6.1975 & 6.1882 & 6.1455 & 6.1706 \\
6 & 0.0536 & 6.3595 & 6.3208 & 6.3447 & 6.3646 \\
7 & 0.0270 & 6.3833 & 6.2950 & 6.3225 & 6.4287 \\
8 & 0.0253 & 6.4122 & 6.3356 & 6.2907 & 6.5096 \\
9 & -0.0104 & 6.5234 & 6.3839 & 5.9705 & 6.6424 \\
10 & 0.0289 & 6.4658 & 6.4140 & 6.0071 & 6.5815 \\
11 & 0.0124 & 6.6338 & 6.5149 & 6.0353 & 6.7757 \\
12 & 0.0152 & 6.7007 & 6.5637 & 6.1959 & 6.8889 \\
13 & 0 & 6.8801 & 6.6513 & 6.4958 & 7.1136 \\
\end{tabular}
\end{ruledtabular}
\end{table}

\subsection{Trivial-topological and semimetal-semiconductor phase transitions}

\begin{figure}
\includegraphics{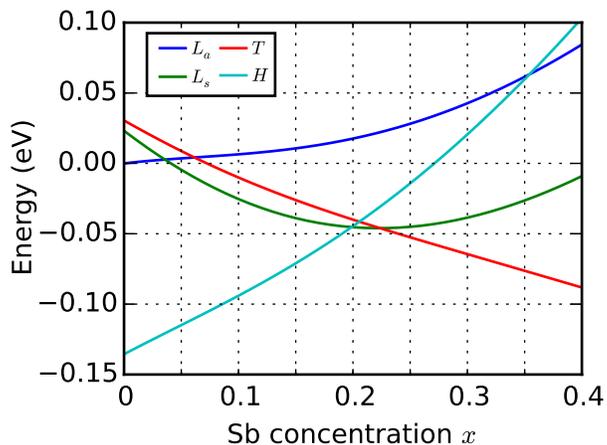}
\caption{\label{fig:statalloy} (Color online) Energies of the four relevant states in Bi$_{1-x}$Sb$_x$ alloys in function of Sb composition $x$. The curves, obtained from a calculation involving the values at Tables~\ref{tab:class} and~\ref{tab:calc}, were subtracted by a constant energy so that the $L_a$ state is at 0 eV on Bi.}
\end{figure}

Having described the computational machinery needed to characterize the Bi$_{1-x}$Sb$_x$ binary alloy, we can predict in a purely theoretical manner the topological and electronic character of the physical system. Results are presented in Fig.~\ref{fig:statalloy}. As discussed previously, the alloy is a (strong) topological insulator (TI) when the bands at the $L$ point are not inverted and the system possesses a band gap.

Schematically, this means the system is TI when the $L_a$ curve in Fig.~\ref{fig:statalloy} is above all other states. The graph must be compared to the diagrams commonly provided in experimental studies\cite{Lenoir1996,Vecchi1977}, and has many peculiar characteristics. First of all, we predict that the TT transition and the first SMSC occur, respectively, at $x\approx0.04$ and $x\approx0.07$, exactly the values reported in Ref.~\onlinecite{Lenoir1996}. The second SMSC, which we find at $x\approx0.36$ is not as accurately described, with measurements placing it at $x\approx0.22$\cite{Lenoir1996}.

Curiously, a VCA calculation\cite{Rose1987} shows a similar behavior, with the TT and first SMSC being well described, and a discrepancy on the second SMSC. The fact that it is possible to arrive, with purely first-principles methods, at a similar conclusion to a calculation performed with semiempirical pseudopotentials is quite astonishing, and also invites some speculation about the physical properties of the alloy in the Sb-rich region responsible for this error, which might be due to strong temperature dependence of the state energies or some sort of strain introduced in the growth process.

Perhaps the most surprising conclusion is that the evolution of the state energies is not linear, as commonly assumed in the literature\cite{Fu2007b,Lenoir1996,Vecchi1977}. To the author's knowledge, the question whether the energies evolve linearly has not yet been carefully assessed experimentally. Our results indicate that there are significant nonlinearities, essential to the description of the TT and SMSC transitions. By introducing a bowing parameter $b_i$ ($i=L_a,L_s,T,H$), one writes

\begin{equation}
E_i(x) = E^{\mathrm{Bi}}_i(1-x) + E^\mathrm{Sb}_ix -b_ix(1-x).
\end{equation}

The bowing parameter may be either constant or composition dependent ($b_i(x)=\alpha_ix+\beta_i$) and are obtained by a least-squares fit of the formula to the GQCA values. They are given in Table~\ref{tab:bowing}, the composition dependent values provide a more accurate fit and are recommended.

\begin{table}
\caption{\label{tab:bowing}Constant ($b$) and composition-dependent ($b(x)=\alpha x + \beta$) bowing parameters (in eV) for all relevant state energies in Bi$_{1-x}$Sb$_x$.}
\begin{ruledtabular}
\begin{tabular}{ccccc}
 & $L_a$ & $L_s$ & $T$ & $H$\\
\hline
$b$ & 0.587 & 0.602 & 0.959 & 0.548 \\
$\alpha$ & 0.680 & 1.014 & 0.306 & 0.705 \\
$\beta$ & -0.185 & -0.824 & 1.307 & -0.314 \\
\end{tabular}
\end{ruledtabular}
\end{table}

\subsection{Influence of strain}

Since the electronic properties of the materials Bi and Sb depend strongly on the strain applied to the crystal lattice, it is expected that one can relate the dependence of the four relevant state energies to the structural properties of the material. Thus, it is possible to understand which type of strain the alloy is subjected to during the growth process by looking at the dependence of its topological properties within our model.

\begin{table}
\caption{\label{tab:deform}Deformation potentials (in eV) for all relevant state energies in Bi and Sb.}
\begin{ruledtabular}
\begin{tabular}{ccccc}
 & $L_a$ & $L_s$ & $T$ & $H$\\
\hline
$a_1$ (Bi) & -11.402 & -11.483 & -12.951 & -13.150 \\
$a_2$ (Bi)& -11.437 & -12.438 & -7.154 & -9.741 \\
$a_1$ (Sb) & -13.486 & -10.946 & -18.175 & -15.180 \\
$a_2$ (Sb)& -12.297 & -10.988 & -9.585 & -10.876 \\
\end{tabular}
\end{ruledtabular}
\end{table}

In order to do that, we evaluate the deformation potentials\cite{Wagner2002} $a_1$ and $a_2$ for both Bi and Sb, and provide them in Table~\ref{tab:deform}. The strain-induced level shifts are given in function of the stress tensor $\hat{\epsilon}$ by the formula

\begin{equation}
E_i(\hat{\epsilon}) = E_{i0} + a^i_1\epsilon_{zz} + a^i_2(\epsilon_{xx}+\epsilon_{yy}).
\end{equation}

With it, it is possible to discuss the effect of biaxial ($\epsilon_{xx}=\epsilon{yy}\neq0$, $\epsilon_{zz}=0$) and uniaxial ($\epsilon_{xx}=\epsilon{yy}=0$, $\epsilon_{zz}\neq0$) on the dependence of state energies. Since the Bi-rich region is well described, we focus mostly on the level shifts caused by strain on Sb. It is assumed, for simplicity, that the level shift depends linearly on Sb composition:

\begin{equation}
\Delta E(\hat{\epsilon}) = x\Delta E^\mathrm{Sb}(\hat{\epsilon})
\end{equation}

\begin{figure}
\includegraphics{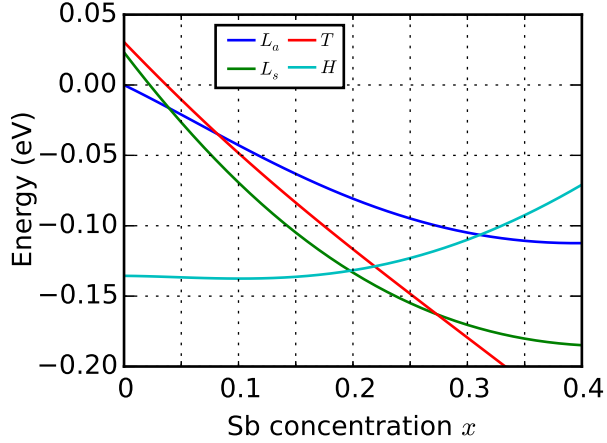}
\caption{\label{fig:biaxial} (Color online) Energies of the four relevant states in Bi$_{1-x}$Sb$_x$ alloys in function of Sb composition $x$ for positive biaxial strain $\epsilon_{xx}=\epsilon_{yy}=0.02$.}
\end{figure}

\begin{figure}
\includegraphics{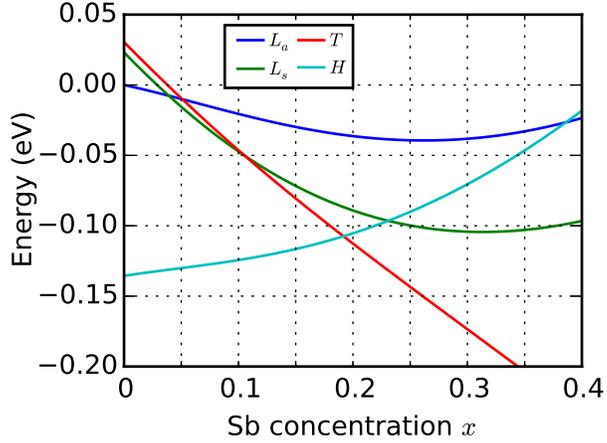}
\caption{\label{fig:uniaxial} (Color online) Energies of the four relevant states in Bi$_{1-x}$Sb$_x$ alloys in function of Sb composition $x$ for positive uniaxial strain $\epsilon_{zz}=0.02$.}
\end{figure}

The results for positive biaxial and uniaxial strain are given, respectively, at Figs.~\ref{fig:biaxial} and~\ref{fig:uniaxial}. It is seen that the effects of the two types of strain are opposite. Increasing biaxial strain causes the second SMSC to occur at lower composition, and raises the energy curve of the $T$ state. Meanwhile, increasing uniaxial strain places the crossing between the $T$ and $L_s$ states at $x\approx0.10$, much closer to the experimentally measured $x=0.09$\cite{Lenoir1996}. However, it also pushes the second SMSC even farther from the measured value at 0.22.

From this, it is possible to infer that, if strain on Sb atoms is responsible for the discrepancy at larger Sb compositions within our model, the type of strain must change drastically as one varies the concentration $x$, being uniaxial at lower composition and biaxial at higher ones. Another alternative is that the strain is of only one type, but changes sign as $x$ is increased.

\section{Conclusions}

In summary, we have described the topological character of Bi$_{1-x}$Sb$_x$ by means of a rigorous statistical model within a quasiparticle-corrected DFT-1/2 calculation. It was shown that the TT and SMSC transitions are reasonably well described, and the energy dependence of the states in function of composition deviates strongly from a linear behavior, presenting large bowings.

Furthermore, the discrepancy in the calculation was related to possible strain introduced in the growth process of the alloy. It was demonstrated that biaxial and uniaxial strain on Sb atoms have opposite effects on the energy levels of the states relevant to the topological phase transitions. These results lead to the conclusion that the only way to achieve the experimentally measured values is by assuming the strain varies drastically at different Sb concentrations.

\begin{acknowledgements}
The author thanks Brett V. Carlson for computational assistance and acknowledges financial support from CAPES (grant no. 88887.116797/2016-00).
\end{acknowledgements}


\begin{thebibliography}{44}
\expandafter\ifx\csname natexlab\endcsname\relax\def\natexlab#1{#1}\fi
\expandafter\ifx\csname bibnamefont\endcsname\relax
  \def\bibnamefont#1{#1}\fi
\expandafter\ifx\csname bibfnamefont\endcsname\relax
  \def\bibfnamefont#1{#1}\fi
\expandafter\ifx\csname citenamefont\endcsname\relax
  \def\citenamefont#1{#1}\fi
\expandafter\ifx\csname url\endcsname\relax
  \def\url#1{\texttt{#1}}\fi
\expandafter\ifx\csname urlprefix\endcsname\relax\def\urlprefix{URL }\fi
\providecommand{\bibinfo}[2]{#2}
\providecommand{\eprint}[2][]{\url{#2}}

\bibitem[{\citenamefont{Smith and Wolfe}(1962)}]{Smith1962}
\bibinfo{author}{\bibfnamefont{G.~E.} \bibnamefont{Smith}} \bibnamefont{and}
  \bibinfo{author}{\bibfnamefont{R.}~\bibnamefont{Wolfe}}, \bibinfo{journal}{J.
  Appl. Phys.} \textbf{\bibinfo{volume}{33}}, \bibinfo{pages}{841}
  (\bibinfo{year}{1962}).

\bibitem[{\citenamefont{Lin et~al.}(2001)\citenamefont{Lin, Cronin, Rabin,
  Ying, and Dresselhaus}}]{Lin2001}
\bibinfo{author}{\bibfnamefont{Y.-M.} \bibnamefont{Lin}},
  \bibinfo{author}{\bibfnamefont{S.~B.} \bibnamefont{Cronin}},
  \bibinfo{author}{\bibfnamefont{O.}~\bibnamefont{Rabin}},
  \bibinfo{author}{\bibfnamefont{J.~Y.} \bibnamefont{Ying}}, \bibnamefont{and}
  \bibinfo{author}{\bibfnamefont{M.~S.} \bibnamefont{Dresselhaus}},
  \bibinfo{journal}{Appl. Phys. Lett.} \textbf{\bibinfo{volume}{79}},
  \bibinfo{pages}{677} (\bibinfo{year}{2001}).

\bibitem[{\citenamefont{Fu et~al.}(2007)\citenamefont{Fu, Kane, and
  Mele}}]{Fu2007a}
\bibinfo{author}{\bibfnamefont{L.}~\bibnamefont{Fu}},
  \bibinfo{author}{\bibfnamefont{C.~L.} \bibnamefont{Kane}}, \bibnamefont{and}
  \bibinfo{author}{\bibfnamefont{E.~J.} \bibnamefont{Mele}},
  \bibinfo{journal}{Phys. Rev. Lett.} \textbf{\bibinfo{volume}{98}},
  \bibinfo{pages}{106803} (\bibinfo{year}{2007}).

\bibitem[{\citenamefont{Kane and Mele}(2005)}]{Kane2005}
\bibinfo{author}{\bibfnamefont{C.~L.} \bibnamefont{Kane}} \bibnamefont{and}
  \bibinfo{author}{\bibfnamefont{E.~J.} \bibnamefont{Mele}},
  \bibinfo{journal}{Phys. Rev. Lett.} \textbf{\bibinfo{volume}{95}},
  \bibinfo{pages}{226801} (\bibinfo{year}{2005}).

\bibitem[{\citenamefont{Hsieh et~al.}(2008)\citenamefont{Hsieh, Qian, Wray,
  Xia, Hor, Cava, and Hasan}}]{Hsieh2008}
\bibinfo{author}{\bibfnamefont{D.}~\bibnamefont{Hsieh}},
  \bibinfo{author}{\bibfnamefont{D.}~\bibnamefont{Qian}},
  \bibinfo{author}{\bibfnamefont{L.}~\bibnamefont{Wray}},
  \bibinfo{author}{\bibfnamefont{Y.}~\bibnamefont{Xia}},
  \bibinfo{author}{\bibfnamefont{Y.~S.} \bibnamefont{Hor}},
  \bibinfo{author}{\bibfnamefont{R.}~\bibnamefont{Cava}}, \bibnamefont{and}
  \bibinfo{author}{\bibfnamefont{M.~Z.} \bibnamefont{Hasan}},
  \bibinfo{journal}{Nature} \textbf{\bibinfo{volume}{452}},
  \bibinfo{pages}{970} (\bibinfo{year}{2008}).

\bibitem[{\citenamefont{Fu and Kane}(2007)}]{Fu2007b}
\bibinfo{author}{\bibfnamefont{L.}~\bibnamefont{Fu}} \bibnamefont{and}
  \bibinfo{author}{\bibfnamefont{C.~L.} \bibnamefont{Kane}},
  \bibinfo{journal}{Phys. Rev. B} \textbf{\bibinfo{volume}{76}},
  \bibinfo{pages}{045302} (\bibinfo{year}{2007}).

\bibitem[{\citenamefont{Wang et~al.}(2016)\citenamefont{Wang, Lang, and
  Kou}}]{Wang2016}
\bibinfo{author}{\bibfnamefont{K.~L.} \bibnamefont{Wang}},
  \bibinfo{author}{\bibfnamefont{M.}~\bibnamefont{Lang}}, \bibnamefont{and}
  \bibinfo{author}{\bibfnamefont{X.}~\bibnamefont{Kou}}, in
  \emph{\bibinfo{booktitle}{Handbook of Spintronics}}, edited by
  \bibinfo{editor}{\bibfnamefont{Y.}~\bibnamefont{Xu}},
  \bibinfo{editor}{\bibfnamefont{D.~D.} \bibnamefont{Awschalom}},
  \bibnamefont{and} \bibinfo{editor}{\bibfnamefont{J.}~\bibnamefont{Nitta}}
  (\bibinfo{publisher}{Springer}, \bibinfo{year}{2016}), pp.
  \bibinfo{pages}{431--462}.

\bibitem[{\citenamefont{Qi et~al.}(2009)\citenamefont{Qi, Hughes, Raghu, and
  Zhang}}]{Qi2009}
\bibinfo{author}{\bibfnamefont{X.-L.} \bibnamefont{Qi}},
  \bibinfo{author}{\bibfnamefont{T.~L.} \bibnamefont{Hughes}},
  \bibinfo{author}{\bibfnamefont{S.}~\bibnamefont{Raghu}}, \bibnamefont{and}
  \bibinfo{author}{\bibfnamefont{S.-C.} \bibnamefont{Zhang}},
  \bibinfo{journal}{Phys. Rev. Lett.} \textbf{\bibinfo{volume}{102}},
  \bibinfo{pages}{187001} (\bibinfo{year}{2009}).

\bibitem[{\citenamefont{Hasan and Kane}(2010)}]{Hasan2010}
\bibinfo{author}{\bibfnamefont{M.~Z.} \bibnamefont{Hasan}} \bibnamefont{and}
  \bibinfo{author}{\bibfnamefont{C.~L.} \bibnamefont{Kane}},
  \bibinfo{journal}{Rev. Mod. Phys.} \textbf{\bibinfo{volume}{82}},
  \bibinfo{pages}{3045} (\bibinfo{year}{2010}).

\bibitem[{\citenamefont{Cho et~al.}(1999)\citenamefont{Cho, DiVenere, Wong,
  Ketterson, and Meyer}}]{Cho1999}
\bibinfo{author}{\bibfnamefont{S.}~\bibnamefont{Cho}},
  \bibinfo{author}{\bibfnamefont{A.}~\bibnamefont{DiVenere}},
  \bibinfo{author}{\bibfnamefont{G.~K.} \bibnamefont{Wong}},
  \bibinfo{author}{\bibfnamefont{J.~B.} \bibnamefont{Ketterson}},
  \bibnamefont{and} \bibinfo{author}{\bibfnamefont{J.~R.} \bibnamefont{Meyer}},
  \bibinfo{journal}{Phys. Rev. B} \textbf{\bibinfo{volume}{59}},
  \bibinfo{pages}{10691} (\bibinfo{year}{1999}).

\bibitem[{\citenamefont{Kosarev}(2007)}]{Kosarev2007}
\bibinfo{author}{\bibfnamefont{V.~V.} \bibnamefont{Kosarev}},
  \bibinfo{journal}{Phys. Solid State} \textbf{\bibinfo{volume}{49}},
  \bibinfo{pages}{2076} (\bibinfo{year}{2007}).

\bibitem[{\citenamefont{Liu and Allen}(1995)}]{Liu1995}
\bibinfo{author}{\bibfnamefont{Y.}~\bibnamefont{Liu}} \bibnamefont{and}
  \bibinfo{author}{\bibfnamefont{R.~E.} \bibnamefont{Allen}},
  \bibinfo{journal}{Phys. Rev. B} \textbf{\bibinfo{volume}{52}},
  \bibinfo{pages}{1566} (\bibinfo{year}{1995}).

\bibitem[{\citenamefont{Teo et~al.}(2008)\citenamefont{Teo, Fu, and
  Kane}}]{Teo2008}
\bibinfo{author}{\bibfnamefont{J.~C.~Y.} \bibnamefont{Teo}},
  \bibinfo{author}{\bibfnamefont{L.}~\bibnamefont{Fu}}, \bibnamefont{and}
  \bibinfo{author}{\bibfnamefont{C.~L.} \bibnamefont{Kane}},
  \bibinfo{journal}{Phys. Rev. B} \textbf{\bibinfo{volume}{78}},
  \bibinfo{pages}{045426} (\bibinfo{year}{2008}).

\bibitem[{\citenamefont{Liu et~al.}(2011)\citenamefont{Liu, Liu, Wu, Duan, Liu,
  and Wu}}]{Liu2011}
\bibinfo{author}{\bibfnamefont{Z.}~\bibnamefont{Liu}},
  \bibinfo{author}{\bibfnamefont{C.-X.} \bibnamefont{Liu}},
  \bibinfo{author}{\bibfnamefont{Y.-S.} \bibnamefont{Wu}},
  \bibinfo{author}{\bibfnamefont{W.-H.} \bibnamefont{Duan}},
  \bibinfo{author}{\bibfnamefont{F.}~\bibnamefont{Liu}}, \bibnamefont{and}
  \bibinfo{author}{\bibfnamefont{J.}~\bibnamefont{Wu}}, \bibinfo{journal}{Phys.
  Rev. Lett.} \textbf{\bibinfo{volume}{107}}, \bibinfo{pages}{136805}
  (\bibinfo{year}{2011}).

\bibitem[{\citenamefont{Aguilera et~al.}(2015)\citenamefont{Aguilera,
  Friedrich, and Bl\"ugel}}]{Aguilera2015}
\bibinfo{author}{\bibfnamefont{I.}~\bibnamefont{Aguilera}},
  \bibinfo{author}{\bibfnamefont{C.}~\bibnamefont{Friedrich}},
  \bibnamefont{and} \bibinfo{author}{\bibfnamefont{S.}~\bibnamefont{Bl\"ugel}},
  \bibinfo{journal}{Phys. Rev. B} \textbf{\bibinfo{volume}{91}},
  \bibinfo{pages}{125129} (\bibinfo{year}{2015}).

\bibitem[{\citenamefont{Falicov and Golin}(1965)}]{Falicov1965}
\bibinfo{author}{\bibfnamefont{L.~M.} \bibnamefont{Falicov}} \bibnamefont{and}
  \bibinfo{author}{\bibfnamefont{S.}~\bibnamefont{Golin}},
  \bibinfo{journal}{Phys. Rev.} \textbf{\bibinfo{volume}{137}},
  \bibinfo{pages}{A871} (\bibinfo{year}{1965}).

\bibitem[{\citenamefont{Falicov and Lin}(1966)}]{Falicov1966}
\bibinfo{author}{\bibfnamefont{L.~M.} \bibnamefont{Falicov}} \bibnamefont{and}
  \bibinfo{author}{\bibfnamefont{P.~J.} \bibnamefont{Lin}},
  \bibinfo{journal}{Phys. Rev.} \textbf{\bibinfo{volume}{141}},
  \bibinfo{pages}{562} (\bibinfo{year}{1966}).

\bibitem[{\citenamefont{Sher et~al.}(1987)\citenamefont{Sher, van Schilfgaarde,
  Chen, and Chen}}]{Sher1987}
\bibinfo{author}{\bibfnamefont{A.}~\bibnamefont{Sher}},
  \bibinfo{author}{\bibfnamefont{M.}~\bibnamefont{van Schilfgaarde}},
  \bibinfo{author}{\bibfnamefont{A.-B.} \bibnamefont{Chen}}, \bibnamefont{and}
  \bibinfo{author}{\bibfnamefont{W.}~\bibnamefont{Chen}},
  \bibinfo{journal}{Phys. Rev. B} \textbf{\bibinfo{volume}{36}},
  \bibinfo{pages}{4279} (\bibinfo{year}{1987}).

\bibitem[{\citenamefont{Freitas
  et~al.}(2016{\natexlab{a}})\citenamefont{Freitas, Furthmüller, Bechstedt,
  Marques, and Teles}}]{Freitas2016a}
\bibinfo{author}{\bibfnamefont{F.~L.} \bibnamefont{Freitas}},
  \bibinfo{author}{\bibfnamefont{J.}~\bibnamefont{Furthmüller}},
  \bibinfo{author}{\bibfnamefont{F.}~\bibnamefont{Bechstedt}},
  \bibinfo{author}{\bibfnamefont{M.}~\bibnamefont{Marques}}, \bibnamefont{and}
  \bibinfo{author}{\bibfnamefont{L.~K.} \bibnamefont{Teles}},
  \bibinfo{journal}{Appl. Phys. Lett.} \textbf{\bibinfo{volume}{108}},
  \bibinfo{pages}{092101} (\bibinfo{year}{2016}{\natexlab{a}}).

\bibitem[{\citenamefont{Freitas
  et~al.}(2016{\natexlab{b}})\citenamefont{Freitas, Marques, and
  Teles}}]{Freitas2016b}
\bibinfo{author}{\bibfnamefont{F.~L.} \bibnamefont{Freitas}},
  \bibinfo{author}{\bibfnamefont{M.}~\bibnamefont{Marques}}, \bibnamefont{and}
  \bibinfo{author}{\bibfnamefont{L.~K.} \bibnamefont{Teles}},
  \bibinfo{journal}{AIP Adv.} \textbf{\bibinfo{volume}{6}},
  \bibinfo{pages}{085308} (\bibinfo{year}{2016}{\natexlab{b}}).

\bibitem[{\citenamefont{Hungerford}(2003)}]{Hungerford2003}
\bibinfo{author}{\bibfnamefont{T.}~\bibnamefont{Hungerford}},
  \emph{\bibinfo{title}{Algebra}}, Graduate Texts in Mathematics
  (\bibinfo{publisher}{Springer New York}, \bibinfo{year}{2003}).

\bibitem[{\citenamefont{Ashcroft and Mermin}(1976)}]{Ashcroft1976}
\bibinfo{author}{\bibfnamefont{N.}~\bibnamefont{Ashcroft}} \bibnamefont{and}
  \bibinfo{author}{\bibfnamefont{N.}~\bibnamefont{Mermin}},
  \emph{\bibinfo{title}{Solid State Physics}} (\bibinfo{publisher}{Holt,
  Rinehart and Winston}, \bibinfo{year}{1976}).

\bibitem[{\citenamefont{Cormen et~al.}(2001)\citenamefont{Cormen, Leiserson,
  Rivest, and Stein}}]{Cormen2001}
\bibinfo{author}{\bibfnamefont{T.}~\bibnamefont{Cormen}},
  \bibinfo{author}{\bibfnamefont{C.}~\bibnamefont{Leiserson}},
  \bibinfo{author}{\bibfnamefont{R.}~\bibnamefont{Rivest}}, \bibnamefont{and}
  \bibinfo{author}{\bibfnamefont{C.}~\bibnamefont{Stein}},
  \emph{\bibinfo{title}{Introduction To Algorithms}} (\bibinfo{publisher}{MIT
  Press}, \bibinfo{year}{2001}).

\bibitem[{\citenamefont{Perdew et~al.}(2008)\citenamefont{Perdew, Ruzsinszky,
  Csonka, Vydrov, Scuseria, Constantin, Zhou, and Burke}}]{Perdew2008}
\bibinfo{author}{\bibfnamefont{J.~P.} \bibnamefont{Perdew}},
  \bibinfo{author}{\bibfnamefont{A.}~\bibnamefont{Ruzsinszky}},
  \bibinfo{author}{\bibfnamefont{G.~I.} \bibnamefont{Csonka}},
  \bibinfo{author}{\bibfnamefont{O.~A.} \bibnamefont{Vydrov}},
  \bibinfo{author}{\bibfnamefont{G.~E.} \bibnamefont{Scuseria}},
  \bibinfo{author}{\bibfnamefont{L.~A.} \bibnamefont{Constantin}},
  \bibinfo{author}{\bibfnamefont{X.}~\bibnamefont{Zhou}}, \bibnamefont{and}
  \bibinfo{author}{\bibfnamefont{K.}~\bibnamefont{Burke}},
  \bibinfo{journal}{Phys. Rev. Lett.} \textbf{\bibinfo{volume}{100}},
  \bibinfo{pages}{136406} (\bibinfo{year}{2008}).

\bibitem[{\citenamefont{Schiferl and Barrett}(1969)}]{Schiferl1969}
\bibinfo{author}{\bibfnamefont{D.}~\bibnamefont{Schiferl}} \bibnamefont{and}
  \bibinfo{author}{\bibfnamefont{C.~S.} \bibnamefont{Barrett}},
  \bibinfo{journal}{J. Appl. Crystallogr.} \textbf{\bibinfo{volume}{2}},
  \bibinfo{pages}{30} (\bibinfo{year}{1969}).

\bibitem[{\citenamefont{Vecchi and Dresselhaus}(1974)}]{Vecchi1974}
\bibinfo{author}{\bibfnamefont{M.~P.} \bibnamefont{Vecchi}} \bibnamefont{and}
  \bibinfo{author}{\bibfnamefont{M.~S.} \bibnamefont{Dresselhaus}},
  \bibinfo{journal}{Phys. Rev. B} \textbf{\bibinfo{volume}{10}},
  \bibinfo{pages}{771} (\bibinfo{year}{1974}).

\bibitem[{\citenamefont{Brandt et~al.}(1982)\citenamefont{Brandt, Hermann,
  Golysheva, Devyatkova, Kusnik, Kraak, and Ponomarev}}]{Brandt1982}
\bibinfo{author}{\bibfnamefont{N.}~\bibnamefont{Brandt}},
  \bibinfo{author}{\bibfnamefont{R.}~\bibnamefont{Hermann}},
  \bibinfo{author}{\bibfnamefont{G.}~\bibnamefont{Golysheva}},
  \bibinfo{author}{\bibfnamefont{L.}~\bibnamefont{Devyatkova}},
  \bibinfo{author}{\bibfnamefont{D.}~\bibnamefont{Kusnik}},
  \bibinfo{author}{\bibfnamefont{W.}~\bibnamefont{Kraak}}, \bibnamefont{and}
  \bibinfo{author}{\bibfnamefont{Y.~G.} \bibnamefont{Ponomarev}},
  \bibinfo{journal}{Zh. Eksp. Teor. Fiz} \textbf{\bibinfo{volume}{83}},
  \bibinfo{pages}{2152} (\bibinfo{year}{1982}).

\bibitem[{\citenamefont{Kohn and Sham}(1965)}]{Kohn1965}
\bibinfo{author}{\bibfnamefont{W.}~\bibnamefont{Kohn}} \bibnamefont{and}
  \bibinfo{author}{\bibfnamefont{L.~J.} \bibnamefont{Sham}},
  \bibinfo{journal}{Phys. Rev.} \textbf{\bibinfo{volume}{140}},
  \bibinfo{pages}{A1133} (\bibinfo{year}{1965}).

\bibitem[{\citenamefont{Aryasetiawan and Gunnarsson}(1998)}]{Aryasetiawan1998}
\bibinfo{author}{\bibfnamefont{F.}~\bibnamefont{Aryasetiawan}}
  \bibnamefont{and}
  \bibinfo{author}{\bibfnamefont{O.}~\bibnamefont{Gunnarsson}},
  \bibinfo{journal}{Rep. Prog. Phys.} \textbf{\bibinfo{volume}{61}},
  \bibinfo{pages}{237} (\bibinfo{year}{1998}).

\bibitem[{\citenamefont{Ferreira et~al.}(2008)\citenamefont{Ferreira, Marques,
  and Teles}}]{Ferreira2008}
\bibinfo{author}{\bibfnamefont{L.~G.} \bibnamefont{Ferreira}},
  \bibinfo{author}{\bibfnamefont{M.}~\bibnamefont{Marques}}, \bibnamefont{and}
  \bibinfo{author}{\bibfnamefont{L.~K.} \bibnamefont{Teles}},
  \bibinfo{journal}{Phys. Rev. B} \textbf{\bibinfo{volume}{78}},
  \bibinfo{pages}{125116} (\bibinfo{year}{2008}).

\bibitem[{\citenamefont{Bl\"ochl}(1994)}]{Blochl1994}
\bibinfo{author}{\bibfnamefont{P.~E.} \bibnamefont{Bl\"ochl}},
  \bibinfo{journal}{Phys. Rev. B} \textbf{\bibinfo{volume}{50}},
  \bibinfo{pages}{17953} (\bibinfo{year}{1994}).

\bibitem[{\citenamefont{Kresse and Furthm\"uller}(1996)}]{Kresse1996a}
\bibinfo{author}{\bibfnamefont{G.}~\bibnamefont{Kresse}} \bibnamefont{and}
  \bibinfo{author}{\bibfnamefont{J.}~\bibnamefont{Furthm\"uller}},
  \bibinfo{journal}{Phys. Rev. B} \textbf{\bibinfo{volume}{54}},
  \bibinfo{pages}{11169} (\bibinfo{year}{1996}).

\bibitem[{\citenamefont{Kresse and Furthmüller}(1996)}]{Kresse1996b}
\bibinfo{author}{\bibfnamefont{G.}~\bibnamefont{Kresse}} \bibnamefont{and}
  \bibinfo{author}{\bibfnamefont{J.}~\bibnamefont{Furthmüller}},
  \bibinfo{journal}{Comp. Mat. Sci.} \textbf{\bibinfo{volume}{6}},
  \bibinfo{pages}{15 } (\bibinfo{year}{1996}).

\bibitem[{\citenamefont{Monkhorst and Pack}(1976)}]{Monkhorst1976}
\bibinfo{author}{\bibfnamefont{H.~J.} \bibnamefont{Monkhorst}}
  \bibnamefont{and} \bibinfo{author}{\bibfnamefont{J.~D.} \bibnamefont{Pack}},
  \bibinfo{journal}{Phys. Rev. B} \textbf{\bibinfo{volume}{13}},
  \bibinfo{pages}{5188} (\bibinfo{year}{1976}).

\bibitem[{\citenamefont{Yazyev et~al.}(2012)\citenamefont{Yazyev, Kioupakis,
  Moore, and Louie}}]{Yazyev2012}
\bibinfo{author}{\bibfnamefont{O.~V.} \bibnamefont{Yazyev}},
  \bibinfo{author}{\bibfnamefont{E.}~\bibnamefont{Kioupakis}},
  \bibinfo{author}{\bibfnamefont{J.~E.} \bibnamefont{Moore}}, \bibnamefont{and}
  \bibinfo{author}{\bibfnamefont{S.~G.} \bibnamefont{Louie}},
  \bibinfo{journal}{Phys. Rev. B} \textbf{\bibinfo{volume}{85}},
  \bibinfo{pages}{161101} (\bibinfo{year}{2012}).

\bibitem[{\citenamefont{K\"ufner and Bechstedt}(2015)}]{Kufner2015}
\bibinfo{author}{\bibfnamefont{S.}~\bibnamefont{K\"ufner}} \bibnamefont{and}
  \bibinfo{author}{\bibfnamefont{F.}~\bibnamefont{Bechstedt}},
  \bibinfo{journal}{Phys. Rev. B} \textbf{\bibinfo{volume}{91}},
  \bibinfo{pages}{035311} (\bibinfo{year}{2015}).

\bibitem[{\citenamefont{Yu et~al.}(2011)\citenamefont{Yu, Qi, Bernevig, Fang,
  and Dai}}]{Yu2011}
\bibinfo{author}{\bibfnamefont{R.}~\bibnamefont{Yu}},
  \bibinfo{author}{\bibfnamefont{X.~L.} \bibnamefont{Qi}},
  \bibinfo{author}{\bibfnamefont{A.}~\bibnamefont{Bernevig}},
  \bibinfo{author}{\bibfnamefont{Z.}~\bibnamefont{Fang}}, \bibnamefont{and}
  \bibinfo{author}{\bibfnamefont{X.}~\bibnamefont{Dai}},
  \bibinfo{journal}{Phys. Rev. B} \textbf{\bibinfo{volume}{84}},
  \bibinfo{pages}{075119} (\bibinfo{year}{2011}).

\bibitem[{\citenamefont{Fruchart and Carpentier}(2013)}]{Fruchart2013}
\bibinfo{author}{\bibfnamefont{M.}~\bibnamefont{Fruchart}} \bibnamefont{and}
  \bibinfo{author}{\bibfnamefont{D.}~\bibnamefont{Carpentier}},
  \bibinfo{journal}{C. R. Phys.} \textbf{\bibinfo{volume}{14}},
  \bibinfo{pages}{779} (\bibinfo{year}{2013}).

\bibitem[{\citenamefont{Drakos and Moore}(1999)}]{VaspManual}
\bibinfo{author}{\bibfnamefont{N.}~\bibnamefont{Drakos}} \bibnamefont{and}
  \bibinfo{author}{\bibfnamefont{R.}~\bibnamefont{Moore}},
  \emph{\bibinfo{title}{Atoms}} (\bibinfo{year}{1999}),
  \urlprefix\url{https://cms.mpi.univie.ac.at/vasp/vasp/Atoms.html}.

\bibitem[{\citenamefont{Popescu and Zunger}(2010)}]{Popescu2010}
\bibinfo{author}{\bibfnamefont{V.}~\bibnamefont{Popescu}} \bibnamefont{and}
  \bibinfo{author}{\bibfnamefont{A.}~\bibnamefont{Zunger}},
  \bibinfo{journal}{Phys. Rev. Lett.} \textbf{\bibinfo{volume}{104}},
  \bibinfo{pages}{236403} (\bibinfo{year}{2010}).

\bibitem[{\citenamefont{Lenoir et~al.}(1996)\citenamefont{Lenoir, Cassart,
  Michenaud, Scherrer, and Scherrer}}]{Lenoir1996}
\bibinfo{author}{\bibfnamefont{B.}~\bibnamefont{Lenoir}},
  \bibinfo{author}{\bibfnamefont{M.}~\bibnamefont{Cassart}},
  \bibinfo{author}{\bibfnamefont{J.-P.} \bibnamefont{Michenaud}},
  \bibinfo{author}{\bibfnamefont{H.}~\bibnamefont{Scherrer}}, \bibnamefont{and}
  \bibinfo{author}{\bibfnamefont{S.}~\bibnamefont{Scherrer}},
  \bibinfo{journal}{J. Phys. Chem. Solids} \textbf{\bibinfo{volume}{57}},
  \bibinfo{pages}{89 } (\bibinfo{year}{1996}).

\bibitem[{\citenamefont{Vecchi et~al.}(1977)\citenamefont{Vecchi, Mendez, and
  Dresselhaus}}]{Vecchi1977}
\bibinfo{author}{\bibfnamefont{M.~P.} \bibnamefont{Vecchi}},
  \bibinfo{author}{\bibfnamefont{E.}~\bibnamefont{Mendez}}, \bibnamefont{and}
  \bibinfo{author}{\bibfnamefont{M.}~\bibnamefont{Dresselhaus}},
  \bibinfo{journal}{Physica B+C} \textbf{\bibinfo{volume}{89}},
  \bibinfo{pages}{150 } (\bibinfo{year}{1977}).

\bibitem[{\citenamefont{Rose and Schuchardt}(1987)}]{Rose1987}
\bibinfo{author}{\bibfnamefont{J.}~\bibnamefont{Rose}} \bibnamefont{and}
  \bibinfo{author}{\bibfnamefont{R.}~\bibnamefont{Schuchardt}},
  \bibinfo{journal}{phys. stat. sol. (b)} \textbf{\bibinfo{volume}{139}},
  \bibinfo{pages}{499} (\bibinfo{year}{1987}).

\bibitem[{\citenamefont{Wagner and Bechstedt}(2002)}]{Wagner2002}
\bibinfo{author}{\bibfnamefont{J.-M.} \bibnamefont{Wagner}} \bibnamefont{and}
  \bibinfo{author}{\bibfnamefont{F.}~\bibnamefont{Bechstedt}},
  \bibinfo{journal}{phys. stat. sol. (b)} \textbf{\bibinfo{volume}{234}},
  \bibinfo{pages}{965} (\bibinfo{year}{2002}).

\end{thebibliography}
\end{document}